\documentclass[12pt]{article}
\usepackage[utf8]{inputenc}
\usepackage{graphicx,xcolor,setspace,geometry}
\usepackage{amsmath}
\usepackage{natbib,authblk,bbm,bm}
\usepackage{multirow}
\usepackage{booktabs}
\usepackage{caption}
\title{A primer on coupled state-switching models\\ for multiple interacting time series}
\author{
Jennifer Pohle$^{1}$\footnote{Corresponding author, \texttt{jennifer.pohle@uni-bielefeld.de}}, 
Roland Langrock$^{1}$, 
Mihaela van der Schaar$^{2,3,4}$,\\
\vspace{-0.75em}
Ruth King$^{3,5}$ and Frants Havmand Jensen$^{6}$\\
\vspace{1em}
\small{
$^1$Bielefeld University, Germany\\
$^2$University of Cambridge, UK\\
$^3$Alan Turing Institute, London, UK\\
$^4$University of California, USA\\
$^5$University of Edinburgh, UK\\
$^6$Woods Hole Oceanographic Institution, USA
}}
\date{}

\begin{document}

\begin{spacing}{1.25}
    \maketitle
\end{spacing}

\begin{spacing}{1.5}

\begin{abstract}
State-switching models such as hidden Markov models or Markov-switching regression models are routinely applied to analyse sequences of observations that are driven by underlying non-observable states. Coupled state-switching models extend these approaches to address the case of multiple observation sequences whose underlying state variables interact. In this paper, we provide an overview of the modelling techniques related to coupling in state-switching models, thereby forming a rich and flexible statistical framework particularly useful for modelling correlated time series. Simulation experiments demonstrate the relevance of being able to account for an asynchronous evolution as well as interactions between the underlying latent processes. The models are further illustrated using two case studies related to a) interactions between a dolphin mother and her calf as inferred from movement data; and b) electronic health record data collected on 696 patients within an intensive care unit.
\end{abstract}

\vspace{0.5em}

%%%%%%%%%%%%%%%%%%%%%%%%%%% 1: Introduction %%%%%%%%%%%%%%%%%%%%%%%%%%%
\section{Introduction}\label{Sec1}

Hidden Markov models (HMMs) are flexible statistical models for sequential data in which the observations are assumed to depend on an underlying latent state process. They have successfully been applied in various areas, starting with speech recognition in the 1970s \citep{dra75} and nowadays including fields such as psychology \citep{vis02}, finance \citep{bul06}, medicine \citep{lan13}, and ecology \citep{mic16}. When modelling multiple observed variables using HMMs, it is usually assumed to have either a) a single state process underlying the observed variables (e.g.\ the speed and tortuosity of an animal's movement are both driven by its behavioural mode), or b) variable-specific but independent state processes (e.g.\ multiple animals separated in space will have independent behavioural modes; \citealp{lan12}). However, there are also scenarios in which neither of these assumptions is valid. For example, multiple individuals may interact due to spatial proximity, the underlying volatilities of different financial markets may affect each other, and body functions may be coupled through physiological mechanisms. In such cases, each process of interest will have its own sequence of underlying states, but the different state processes are \textit{coupled}.

Coupled hidden Markov models (CHMMs) extend the basic HMM framework by assuming distinct but correlated state sequences that underlie the observed variables, hence ``coupling'' the state processes. Since their first appearance in \citet{bra97a}, they have been further developed and applied for example to classify electroencephalography data \citep{mic14}, to model interactions of suspects in forensics \citep{bre06}, and to detect bradycardia events from electrocardiography data \citep{gha16}. CHMMs can be considered as established tools within the engineering literature, where they are commonly applied in classification tasks, e.g.\ emotion recognition from audio-visual signals \citep{lin12}, or gesture recognition from hand tracking data \citep{bra97b}. As a full probabilistic model for sequential data, CHMMs can however also be useful for other inferential purposes, including forecasting future observations as well as general inference on the data-generating process.  

In this work, we argue that the full potential of CHMMs for such statistical modelling challenges to date has not been recognised, as evidenced by the fact that these models have only very rarely been used in such a context; some notable exceptions are \citet{she13}, \citet{joh16}, and \citet{tou19}. We set out to fill this gap, by introducing the CHMM formulation, in particular discussing the various simplifying assumptions that one may or may not want to make, and by presenting inferential tools available for CHMMs. Furthermore, we discuss the inclusion of covariates and introduce a coupled Markov-switching regression (CMSR) model which allows the observed variables to depend on covariates. Simulation studies are used to highlight practical issues that are relevant when modelling multiple interacting processes, thereby showcasing the potential benefits of the CHMM framework compared to more basic model formulations. Finally, we illustrate the practical use of CHMMs in two case studies. First, we consider a simple CHMM for studying the behaviour of a dolphin mother and calf pair. Second, we apply a CMSR model to electronic health record data collected by the University of California in Los Angeles (UCLA) to model the evolution of important vital signs over time, controlling for age and sex of the patients.

%%%%%%%%%%%%%%%%%%%%%%%%%%% 2: HMMs and CHMMs %%%%%%%%%%%%%%%%%%%%%%%%%%%
\section{Hidden and coupled hidden Markov models}\label{Sec2} 

\subsection{Hidden Markov models}\label{Sec2.1}

\subsubsection{Basic model formulation for univariate time series}\label{Sec2.1.1}

An HMM is a doubly stochastic process comprising an observable time series $\{Y_t\}_{t=1}^{T}$ and an underlying latent state sequence $\{S_t\}_{t=1}^{T}$. In the basic model formulation, the state sequence is a first-order Markov chain, i.e.\ 
$$\Pr(S_t=s_{t}|S_{t-1}=s_{t-1},\ldots,S_{1}=s_1)=\Pr(S_{t}=s_{t}|S_{t-1}=s_{t-1}),$$
with $S_{t} \in \{1,2,\ldots,N\}$. To simplify notation, we will abbreviate expressions such as the one above to $\Pr(S_t|S_{t-1},\ldots,S_{1})=\Pr(S_{t}|S_{t-1})$.  
The state transition probabilities are summarised in the transition probability matrix (TPM) $\boldsymbol{\Gamma}=(\gamma_{ij})$, with $\gamma_{ij}=\Pr(S_{t}=j|S_{t-1}=i)$, $i,j=1,\ldots,N$. We assume the Markov chain to be homogeneous and stationary, unless explicitly stated otherwise. The initial distribution, $\bm{\delta}=\bigl(\Pr(S_{1}=1) , \ldots , \Pr(S_{1}=N)\bigr)$, then is the solution to $\bm{\delta}\bm{\Gamma}=\bm{\delta}$ subject to $\sum_{i=1}^N \delta_i =1$. Given the state at time $t$, the observation at time $t$ is assumed to be conditionally independent of past observations and states, 
$$f(Y_{t}|Y_{t-1},\ldots,Y_1,S_{t},\ldots,S_1)=f(Y_{t}|S_{t})=f_{S_{t}}(Y_{t}),$$ where $f$ is either a probability or a density function. At each time $t$, the observation $Y_{t}$ is thus generated by one of $N$ state-dependent distributions, as selected by the state active at time $t$. In its basic form, an HMM is hence fully characterised by the TPM, $\boldsymbol{\Gamma}$, and the state-dependent distributions, $f_{i}(Y_{t})$, $i=1,\ldots,N$.

\subsubsection{Inference for hidden Markov models}\label{Sec2.1.2}

For a given time series $y_1,\ldots,y_T$, the HMM likelihood can concisely be written as
\vspace{-2em}
\begin{align*}
\mathcal{L} = \bm{\delta} \mathbf{P_{1}} \bm{\Gamma} \mathbf{P_{2}} \cdots \bm{\Gamma} \mathbf{P_{T}} \bm{1}', 
\end{align*}
where $\mathbf{P}_t$ denotes an $N \times N$ diagonal matrix with entries $f_{i}(y_{t})$, $i=1,\ldots,N$, and $\bm{1}$ is an $N$-dimensional row vector of ones. The matrix product expression corresponds to a recursive calculation of the likelihood using the forward algorithm, which comes at a computational cost of order $\mathcal{O}(TN^{2})$. Numerical optimisation routines such as Newton-type procedures, or alternatively the expectation-maximisation (EM) algorithm, can be used to find the maximum likelihood estimate. In a Bayesian estimation framework, Markov chain Monte Carlo methods can be used to sample from the posterior distribution of the parameters \citep{ryd08}, again making use of recursive techniques such as the forward algorithm. The Viterbi algorithm can be applied for global state decoding, i.e.\ to derive the most likely state sequence under the model, given the data. Alternatively, the states can be locally decoded based on maximising the conditional state probabilities at each time point $t=1,\ldots,T$ \citep{zuc16}.

\subsubsection{HMMs for multivariate time series}\label{Sec2.1.3}

We now consider multivariate time series $\{\bm{Y}_{t}\}_{t=1}^{T}$, with $\bm{Y}_{t}=(Y^{(1)}_{t},\ldots,Y^{(M)}_{t})$. In this case, the state-dependent distributions within the HMM are multivariate, e.g.\ $M$--dimensional multivariate normal distributions. However, in practice the $M$ variables observed often have different scales of measurement --- e.g.\ positive continuous, proportion, count, or binary --- rendering it difficult to formulate a suitable joint distribution. Thus, often a third simplifying assumption is made, namely that given the current state $S_{t}$, all $M$ variables are conditionally independent of each other: $f(\bm{Y}_t|S_{t})=\prod_{m=1}^{M} f(Y^{(m)}_{t}|S_{t})$. Under this contemporaneous conditional independence assumption, a suitable class of univariate distributions is chosen separately for each of the $M$ variables.

Irrespective of the specific dependence assumption made for the observed process, a conventional multivariate HMM assumes the $M$ observed time series to be driven by a \textit{single} underlying state sequence. As a consequence, the $M$ variables evolve in lockstep regarding underlying state switches \citep{bra97a}. This will often be a natural assumption, e.g.\ when modelling the movement of an individual animal, where it is common to model observed step lengths and turning angles using a bivariate HMM. In those instances, the states are proxies for the behavioural modes of the animal considered, and thus a change of this mode would be reflected in both speed and tortuosity of movement (see, for example, \citealt{mic16}). Likewise, in financial time series modelling, the model state might be a proxy for the nervousness of the market, with a single corresponding quantity usually sufficient to capture the volatility of multiple share return series being modelled \citep{mar19}. In contrast, if the $M$ variables considered were to evolve completely independently of each other, then it would be adequate to simply fit univariate HMMs separately to each of the $M$ time series. 

For multivariate time series, models assuming a single underlying state sequence and those assuming multiple independent state sequences thus constitute the two extremes in terms of the state-driven dependence structure between variables. We focus on scenarios that fit neither of these two extremes, and instead are such that each of the different variables observed depends on its own underlying state variable, but such that the state variables interact and influence each other. For instance, we consider systems where the state switches often, but not always, occur at the same time. This becomes relevant when modelling interacting individuals or variables. For such a dependence structure, we require HMM formulations for $M$--dimensional time series, with $M$ underlying state variables, such that those $M$ state processes are correlated with each other. Such a structure is provided by coupled hidden Markov models (CHMMs).

\subsection{Coupled hidden Markov models}\label{Sec2.2}

Consider $M$ distinct time series $\{ Y^{(m)}_t \}_{t=1}^{T}$, each depending on an underlying state sequence $\{S_t^{(m)}\}_{t=1}^{T}$, $m=1,\ldots,M$. For notational simplicity we restrict the presentation to the case where each of the $M$ observed processes is univariate, but the extension to multivariate processes is straightforward. CHMMs link the different time series via the state process by allowing the underlying states $S_t^{(m)}$ to interact: in addition to assuming that ${S}_{t-1}^{(m)}$ affects ${S}_{t}^{(m)}$, we allow also ${S}_{t-1}^{(n)}$ to affect ${S}_{t}^{(m)}$ for $n \neq m$. The dependence structure between the state variables is thus reflected in the transition probabilities of the CHMM. The observed variables are again assumed to be conditionally independent given the states.

Next we discuss possible assumptions regarding the exact dependence structure of the state processes within a CHMM, which differ in terms of their flexibility and hence the dimensionality of the parameter space (i.e.\ model complexity). To simplify notation, we assume the state space for each state variable to be of the same dimension, $N$, i.e.\ $|\mathcal{S}^{(m)}|=N$ for $m=1,\ldots,M$; the extension to the more general case is straightforward.

\subsubsection{Cartesian product model}\label{Sec2.2.1}

\begin{figure}[!tb]
\centering
\includegraphics[width=0.7\textwidth]{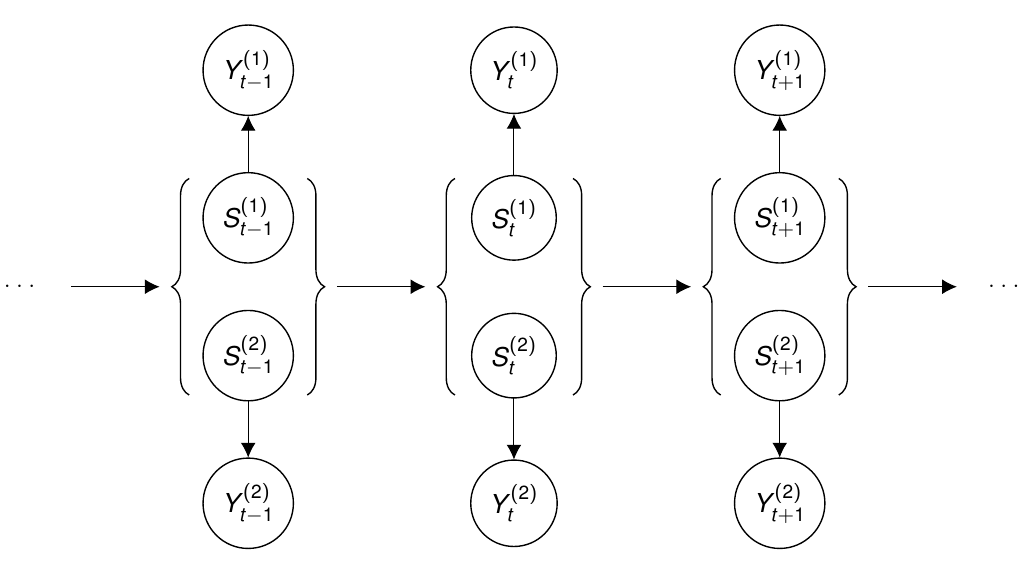}
\caption{Dependence structure of the Cartesian product CHMM with $M=2$ distinct time series.}\label{fig:Cartesian_CHMM}
\end{figure}

Instead of modelling each state variable $S_{t}^{(m)}$ separately, they can be summarised in the $M$-dimensional state vector $\bm{S}_{t}=(S^{(1)}_{t},\ldots,S^{(M)}_{t})$. The CHMM can then be defined as an HMM with the multivariate state sequence $\{\bm{S}_t\}_{t=1}^T$. The corresponding state space $\bm{\mathcal{S}}$ is built by the Cartesian product of all individual state spaces, i.e.\ $\bm{\mathcal{S}}= \mathcal{S}^{(1)} \times \ldots \times \mathcal{S}^{(M)}$, with $|\bm{\mathcal{S}}|=N^{M}$ and the transition probabilities then referring to the state vectors, i.e.\ $\Pr(\bm{S}_{t}|\bm{S}_{t-1}) =\Pr \bigl( (S_t^{(1)},\ldots,S_t^{(M)}) | (S_{t-1}^{(1)},\ldots,S_{t-1}^{(M)}) \bigr)$ (see Figure \ref{fig:Cartesian_CHMM} for an illustration for the case $M=2$). This model formulation is attractive because there is no need to develop new estimation and inference methods: all techniques available for basic HMMs can easily be transferred. 
The Cartesian product formulation comprises two important special cases: 1) independent state processes, corresponding to separately fitting HMMs to each of the M sequences, and 2) multiple observed variables that depend on only a single state sequence, i.e.\ a multivariate HMM. Importantly, it additionally captures the dependence structures in-between these extreme situations. However, this flexibility is associated with a number of parameters that is exponential in the number of state variables $M$, as the TPM is of dimension $N^{M} \times N^{M}$. The computational cost thus is high even for moderate $M$ and $N$. For instance, for $M=3$ and $N=3$, we would have $|\mathcal{S}|=3^{3}=27$ and a TPM of dimension $27 \times 27$ (with only $702$ of the $729$ entries to be estimated, due to the row constraints). This high number of parameters will often lead to numerical problems in the optimisation (e.g.\ local maxima) and may raise the risk of overfitting.

We note here that the use of the label ``coupled HMM'' is in fact not consistent in the existing literature, and that the Cartesian product model is not always regarded as a CHMM (see, for example, \citealp{bra97a}, \citealp{bra97b}, \citealp{nef02}). Other authors use the Cartesian product formulation as a convenient framework for estimation (see, for example, \citealt{rez00,gho17}). In this contribution, the label CHMM refers to all models that couple several HMMs via the state process, and we regard the Cartesian product model as one way to specify such a CHMM.

\subsubsection{CHMM with contemporaneous conditional independence assumption}\label{sec2.2.2}

\begin{figure}[!t]
\centering
\includegraphics[width=0.7\textwidth]{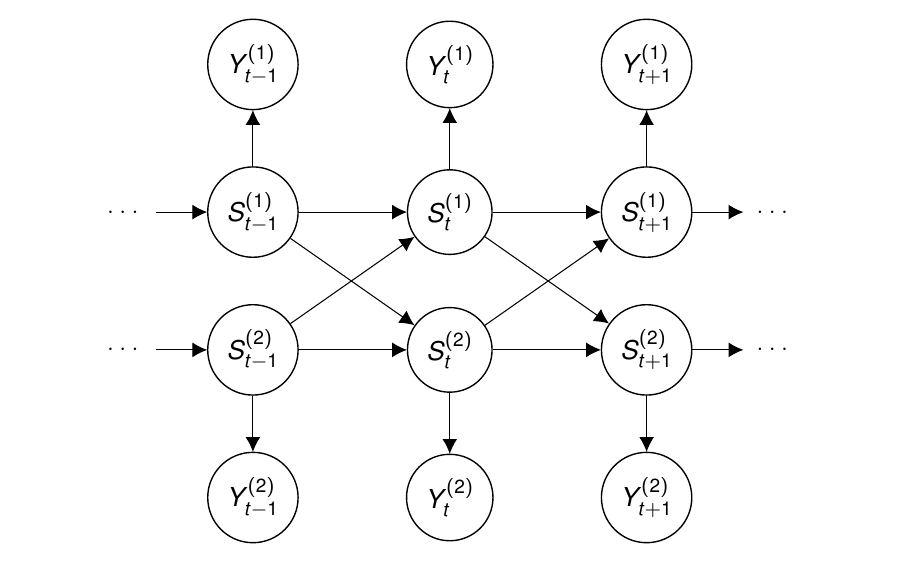}
\caption{CHMM structure with contemporaneous conditional independence assumption for $M=2$ time series.}\label{fig:CI_CHMM}
\end{figure}

The Cartesian product model contains instantaneous correlations between the states, i.e.\ the transition probabilities $\Pr(\bm{S}_{t}|\bm{S}_{t-1})$ cannot be factorised into simpler expressions. Alternatively, the state variables $S^{(m)}_{t}$, $m=1,\ldots,M$, can be assumed to be contemporaneously conditionally independent given the state vector $\bm{S}_{t-1}$: 
$$\Pr(\bm{S}_{t}|\bm{S}_{t-1})=\prod_{m=1}^{M} \Pr(S^{(m)}_{t}|\bm{S}_{t-1});$$
see Figure \ref{fig:CI_CHMM}. 

This model formulation involves $MN^{M+1}$ transition probabilities describing the state dynamics (e.g.\ for $M=3$ and $N=3$, this results in $243$ transition probabilities, $162$ of them to be estimated due to sum constraints). Naturally, this assumption reduces the flexibility of the model: for example, it cannot accommodate patterns where the $M$ state variables tend to switch states simultaneously. For parameter estimation, this CHMM formulation can be converted into a Cartesian product CHMM, thereby opening up the way for all standard HMM machinery. The resulting model would again have a state space of dimension $N^M$, but with restrictions on the transition probabilities due to the states' contemporaneous conditional independence. In fact, the conversion into a Cartesian product formulation can be used to estimate the parameters of \textit{all} CHMM formulations that define a valid probabilistic model on the state vectors.

\subsubsection{CHMMs with explicit modelling of variable-to-variable effects}\label{sec2.2.3}

In the CHMM representations discussed above, there is no parameter explicitly representing direct variable-to-variable effects, which makes interpretation difficult \citep{bra97a}. \citet{sau99} offer a remedy to this caveat by combining the contemporaneous conditional independence assumption with a mixture representation for the marginal transition probabilities:
\begin{align*}
\Pr(\bm{S}_{t}|\bm{S}_{t-1})&=\prod_{m=1}^{M} \Pr(S^{(m)}_{t}|\bm{S}_{t-1}),\\
\text{where }\Pr(S^{(m)}_{t}|\bm{S}_{t-1})&=\sum_{n=1}^{M} w^{(m)}(n) \Pr(S^{(m)}_{t}|S^{(n)}_{t-1}),  
\end{align*}
with $0 \leq w^{(m)}(n) \leq 1$ and $\sum_{n=1}^{M} w^{(m)}(n)=1$. The mixture weight $w^{(m)}(n)$ here reflects the strength of the effect of state $S_{t-1}^{(n)}$ on $S_{t}^{(m)}$ --- independent state processes would result in $w^{(m)}(m)=1$ for all $m=1,\ldots,M$, and $w^{(m)}(n)=0$ for all
$n \neq m$. This model is similar in spirit to the mixture transition duration higher-order Markov chain model suggested by \citet{raf85}. It involves $M^2N^2$ marginal transition probabilities describing the interactions in the state processes, in addition to $M^2$ weights (e.g.\ for $M=3$ and $N=3$, this results in $90$ parameters, $60$ of them to be estimated due to sum constraints). While much reduced in terms of its complexity, we found this model to be numerically unstable as the weights $w^{(m)}(n)$ are often estimated very close to zero for $n\neq m$. Estimation and further inference can again be conducted based on a Cartesian product representation, or alternatively, to avoid the associated large state space, using a bespoke EM algorithm \citep{sau99}.

The CHMM originally proposed by \citet{bra97a} is described by a factorisation based on contemporaneously conditionally independent state variables: $$\Pr(S^{(m)}_{t}|\bm{S}_{t-1})=\prod_{n=1}^{M} \Pr(S^{(m)}_{t}|S^{(n)}_{t-1}).$$ 
This parameterisation, which reduces the number of transition parameters to $M^{2} N^{2}$, may appear intuitive, but it does not yield properly defined transition probabilities for the state vector $\bm{S_{t}}$ as it does not guarantee that $\sum_{\bm{S_t}} \Pr(\bm{S_{t}}|\bm{S_{t-1}}) = 1$. As a consequence, there is no unique conversion to the Cartesian product model, and it is difficult to obtain valid inference. Estimation can be carried out using a (non-unique) mapping to the Cartesian product formulation \citep{bra97b}, by applying a tailored EM algorithm \citep{gha16}, or based on a variational learning approach \citep{bra97a}.

In a Bayesian framework, \citet{she13} propose to directly model the influence of state $S_{t-1}^{(n)}$ on $S_{t}^{(m)}$, $n \neq m$, by using it as a covariate for the state transition probabilities $\Pr(S^{(m)}_{t}|S^{(m)}_{t-1})$. This is possible only as the complete state sequences are drawn within a Gibbs sampler, and the latent states then treated as if they were known within the posterior conditional distribution. Furthermore, in the applied setting described in \citet{she13}, namely the modelling of interactions between diseases in a host, the relation between the states and the observations is deterministic and the structure of the TPMs is known, which greatly facilitates model building and estimation. 

\subsection{Coupled Markov-switching regression}\label{Sec2.3}

We now turn to models which account for the influence of covariates. For example, the transition probabilities of the state process of an HMM can be expressed as a function of covariates using an appropriate link function such as the multinomial logit \citep{zuc16}. While this approach can in principle be applied to CHMMs, it will often be infeasible as even a basic CHMM typically involves a high number of transition probabilities, such that model complexity can be prohibitive. The incorporation of covariates into the observation process --- often referred to as Markov-switching regression (MSR; \citealp{lan17}) --- is more promising for the CHMM setting. MSR models were first introduced for econometric time series, in which case they can be used, for example, to investigate if covariate effects differ between periods of high and low economic growth, respectively \citep{ham08}. The MSR framework can be transferred to the CHMM setting by relating the $M$ observed variables to (variable-specific) covariates, for example as follows:
$$ Y^{(m)}_{t} \,|\, S_t^{(m)}= i \; \sim \; \mathcal{N}(\mu_{m,i,t},\sigma^{2}_{m,i}), $$
$$ \mu_{m,i,t}  = \beta_{0,m,i} + \beta_{1,m,i} \cdot x_{1,t}^{(m)} + \ldots + \beta_{p,m,i} \cdot x_{p,t}^{(m)}, $$
$m=1,\ldots,M$, $t=1,\ldots,T$. In this example model, each of the $M$ variables is conditionally normally distributed, with state- and variable-specific (constant) variance and a state- and variable-specific linear predictor determining the mean. In combining CHMMs and MSR models, this coupled Markov-switching regression (CMSR) model takes into account possible interactions in the state processes underlying the $M$ observed variables, but also the influences of covariates on the observation process. For parameter estimation, the state- and covariate-dependent observation distributions can simply be plugged into the HMM-likelihood function, such that once again the basic HMM machinery remains  applicable. The example model given above can easily be generalised to allow for other distributional families for the observed variables (cf.\ \citealt{lan17}).

%
%%%%%%%%%%%%%%%%%%%%%%%%%%% 3: Simulation %%%%%%%%%%%%%%%%%%%%%%%%%%%
%
\section{Simulation study}\label{Sec3}

We provide simulation experiments to illustrate the consequences of neglecting or misspecifying the dependence structure in the state process. More specifically, we simulate data from a CHMM as the true data-generating process --- i.e.\ multiple time series with interacting underlying state processes --- and demonstrate the consequences of either completely neglecting the interaction (by fitting separate univariate HMMs) or incorrectly assuming full synchronicity (by fitting a multivariate HMM). 

The data-generating process we consider is a Cartesian product CHMM with $M=2$ observed variables and $N=2$ states per variable. The variables  $Y_{t}^{(1)}$ and $Y_{t}^{(2)}$ are thus driven by the underlying bivariate state sequence $\bm{S}_{t}=(S_{t}^{(1)},S_{t}^{(2)})$ with $S_t^{(1)},S_t^{(2)} \in \{1,2\}$, such that the Cartesian product state space is of dimension $|\bm{\mathcal{S}}|=4$. To simplify notation, we fix the order of the states to $(1,1)$ (state 1 of the process $\bm{S}_{t}$), $(1,2)$ (state 2), $(2,1)$ (state 3) and $(2,2)$ (state 4), and refer to this order when defining the TPM and the corresponding stationary distribution. The TPM is chosen such that the random variables evolve synchronously most of the time, i.e.\ the model is only slightly different from a multivariate HMM:
$$\Gamma=\begin{pmatrix}
0.90 & 0.02 & 0.02 & 0.06 \\ 
0.09 & 0.80 & 0.02 & 0.09 \\ 
0.09 & 0.02 & 0.80 & 0.09 \\ 
0.06 & 0.02 & 0.02 & 0.90 \\ 
\end{pmatrix}.$$
The corresponding stationary distribution, $\delta=(0.41, 0.09, 0.09, 0.41)$, indicates that the process is in either of the two states corresponding to synchronicity, i.e.\ $(1,1)$ or $(2,2)$, $82\%$ of the time. For the state-dependent distributions, we assume
$$   Y_{t}^{(1)} \, | \, S_{t}^{(1)} \sim
\begin{cases}
 \mathcal{N}(2,1.5)  & \text{if } S_{t}^{(1)}=1 \\
 \mathcal{N}(6,1.5)  & \text{if } S_{t}^{(1)}=2 \\
\end{cases}, \quad 
Y_{t}^{(2)} \, | \, S_{t}^{(2)} \sim
\begin{cases}
 \mathcal{N}(2,1.5)  & \text{if } S_{t}^{(2)}=1 \\
 \mathcal{N}(5,1.5)  & \text{if } S_{t}^{(2)}=2 \\
\end{cases} 
$$ 
From the CHMM described above, we generate a training data set of size $T=1000$, and an additional test set comprising 100 observations. The following models are fitted to the simulated data: two separate univariate 2-state HMMs, a multivariate 2-state HMM, and a $2\times2$ Cartesian product CHMM, in each case with state-dependent normal distributions. All models are fitted using numerical maximisation of the likelihood. Subsequently, we compare the true and estimated parameters of the state-dependent distributions (estimation accuracy, Section \ref{Sec3.1}), the number of correctly decoded states based on the Viterbi algorithm (classification performance, Section \ref{Sec3.2}), and the conditional likelihood of the test set given the training data (forecasting performance, Section \ref{Sec3.3}). We repeat these steps $1000$ times and compare the results across simulation runs.

\subsection{Estimation accuracy}\label{Sec3.1}

\begin{figure}[t!]
    \centering
    \includegraphics[width=0.7\textwidth]{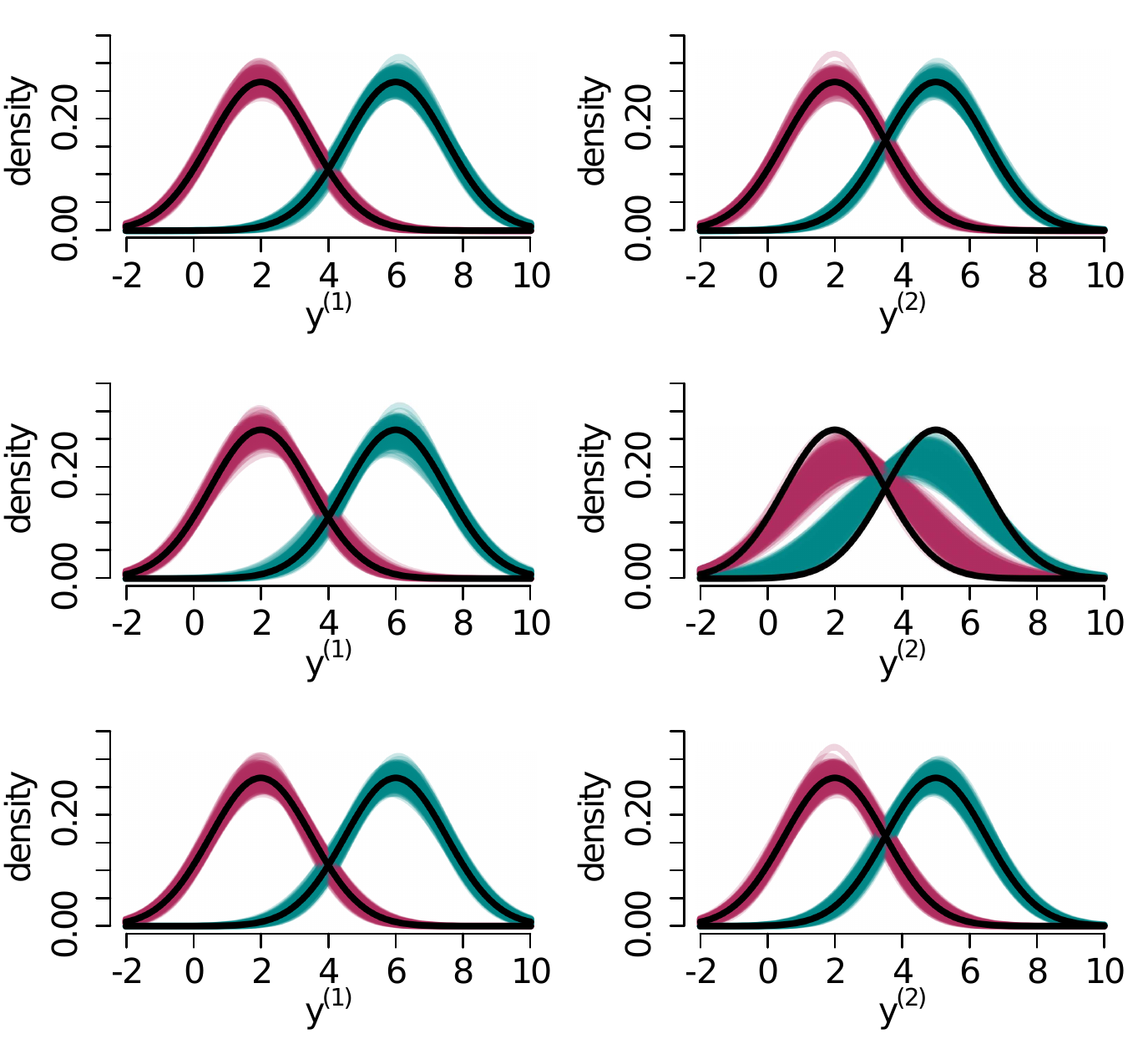}
    \caption{Estimated state-dependent densities obtained in 1000 simulation runs. The upper panel displays the results of the fitted CHMMs, the middle panel corresponds to the multivariate HMMs, and the bottom panel to the estimated univariate HMMs. The black lines show the true underlying densities.}
    \label{fig:sim9_7}
\end{figure}
Figure \ref{fig:sim9_7} displays the state-dependent densities as obtained in the 1000 runs, for each of the three model formulations considered. Under the correct CHMM specification, but also under the incorrect model specification using two separate univariate HMMs, the true state-dependent densities were generally well recovered in the estimation. In other words, even when neglecting the correlation of the two state processes the estimation is fairly accurate at the level of the observation process. However, the situation is fundamentally different when the correlation of the two state sequences is effectively \textit{overestimated}, i.e.\ when using the multivariate HMM formulation, which amounts to assuming the state processes to be completely synchronous. Whenever the simulated state variables $S_{t}^{(1)}$ and $S_{t}^{(2)}$ differ, the multivariate HMM with its single underlying state process cannot correctly identify the state combination anymore --- it effectively distinguishes the pairs $(1,1)$ and $(2,2)$. At those instances, the implicit state allocation is dominated by the $Y_t^{(1)}$ process with its more clearly distinct state-dependent distributions. As a consequence, true state pairs $(1,2)$ and $(2,1)$ are effectively modelled as $(1,1)$ and $(2,2)$ pairs, respectively, such that the estimators of the state-dependent distributions of the $Y_t^{(2)}$ process are heavily biased (towards a middle ground).

\subsection{Classification}\label{Sec3.2}

\begin{table}[t!]
\centering
\begin{tabular}{l|ccc}
\toprule
data set & CHMM & multi. HMM & uni. HMMs  \\ \midrule
training data set & 5.7 & 19.7 & 8.1 \\
test data set & 6.0 & 19.7 & 8.3 \\
\bottomrule
\end{tabular}
\caption{Average percentage of falsely decoded states in the Viterbi sequence.}
\label{tab:sim9_7decoding}
\end{table}

The comparison of the classification performance is based on the globally decoded Viterbi state sequences as obtained for both the training and test data, respectively. Table \ref{tab:sim9_7decoding} displays the average percentage of falsely decoded states across all simulation runs under the univariate, multivariate and CHMMs, respectively. The multivariate HMM has the largest classification error as it cannot correctly identify the state pair if $S^{(1)}_t \neq S^{(2)}_t$. The CHMM outperforms the univariate HMMs as the latter do not take into account the interaction dynamics between the two state processes, which help to inform the decoding.

\subsection{Forecasting performance}\label{Sec3.3}
To compare the forecasting performance, we consider the conditional log-likelihood of the test set given the training data, $\mathcal{L}(\bm{Y}_{test},\hat{\boldsymbol{\theta}}|\bm{Y}_{training})$. The CHMM had the largest conditional log-likelihood in 85.4\% of all runs (this number increases to $99.8\%$ when increasing the sample size of the training set to $5000$ and the size of the test set to $500$). 

In summary, our simulations show that misspecifications of the dependence structure in the state process have various undesirable consequences. Erroneously mistaking two separate, highly correlated state sequences for a single state sequence led to substantially biased estimators, a high classification error and poor forecasting performance. Distinguishing two such state sequences but failing to account for their correlation negatively affected the forecasting and classification performance.

%
%%%%%%%%%%%%%%%%%%%%%%%%%%% 4: Case studies %%%%%%%%%%%%%%%%%%%%%%%%%%%
%
\section{Case studies}\label{Sec4}

We illustrate the application of CHMMs in two case studies. First, we analyse movements of a dolphin mother and its calf using a Cartesian product CHMM. Subsequently, we apply a CMSR model to data on vital signs of patients hospitalised in the intensive care unit (ICU), controlling for sex and age. Parameters were estimated via numerical likelihood maximisation using the R function \texttt{nlm}.

\subsection{Movements of dolphin mother and calf}\label{Sec4.1}

HMMs are routinely used to analyse animal movement data, with the model's state process interpreted as a proxy for an animal's behavioural modes (e.g.\ resting, foraging or relocating) determining the observed movement patterns \citep{lan12}. Here we consider movement data from a bottlenose dolphin mother and calf pair which was simultaneously tagged with 3D accelerometers and magnetometers for ${\sim}18$ hours. Our analysis focuses on the tortuosity of the movement across $10$-second intervals, i.e.\ a measure of how tortuous the dead-reckoned track of the animal is. This results in $T=6546$ tortuosity observations per animal. The values lie in the interval $[0,1)$ with $0$ corresponding to straight-line movement.

It is certain that the two animals interact, i.e.\ that the behaviour of mother and calf influence each other. To account for these interactions, instead of fitting two univariate HMMs separately to both individuals, we consider CHMMs within which the two animals' separate behavioural state sequences are correlated. To avoid restrictive assumptions regarding the interaction, we use a Cartesian product CHMM with bivariate state vectors --- indeed the AIC favoured this ``full'' CHMM over the alternative model formulations that involve more restrictive assumptions (results not shown). Tortuosity was modelled using state-dependent beta distributions. To avoid additional parameters corresponding to point masses on zero, the observed zeros ($2.5\%$ for the mother, $0.2\%$ for the calf) were shifted by very small positive random numbers. We expect that tortuosity in general might reflect multiple different behavioural regimes, from directed resting and travel behaviours to more tortuous back-and-forth scanning movements during biosonar-based foraging, to high tortuosity circling and rapid turning behaviours in connection with prey capture. Thus, for each of the two individuals we considered $N=3$ states.

\begin{figure}[t!]
    \centering
    \includegraphics[width=0.6\textwidth]{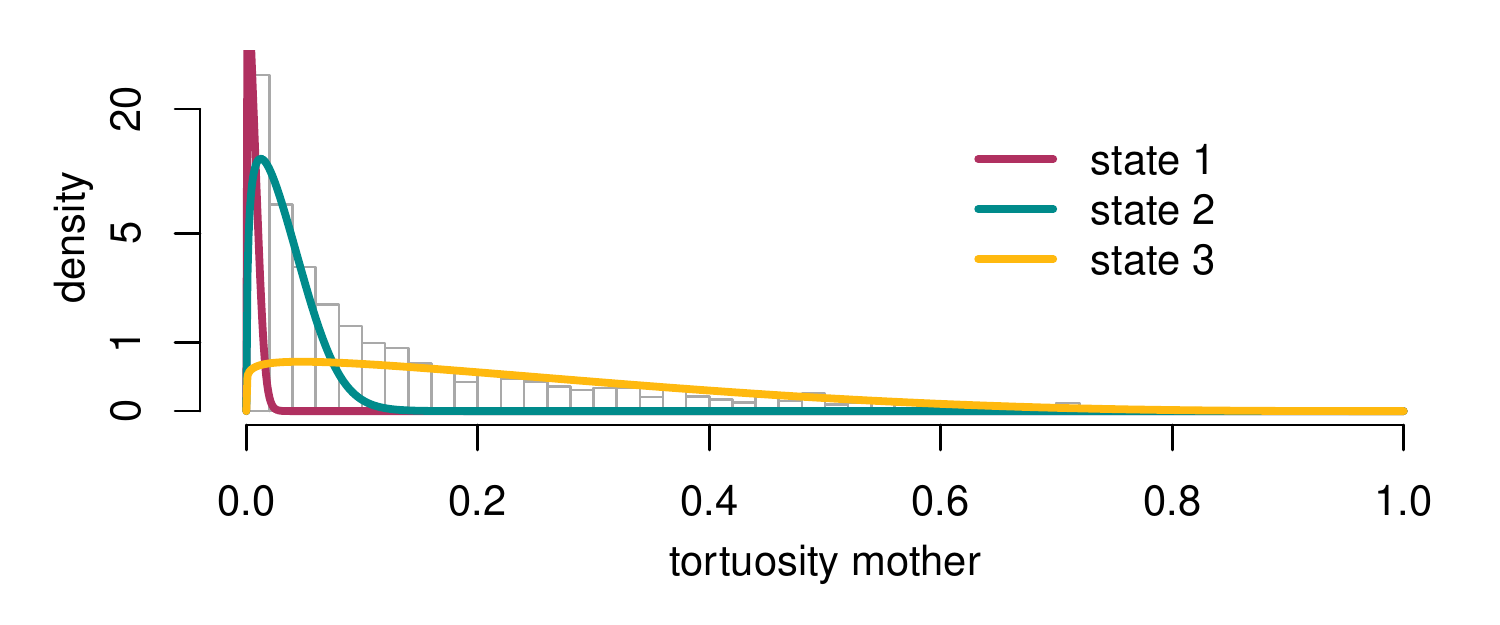}
    
    \vspace{-1em}
    
    \includegraphics[width=0.6\textwidth]{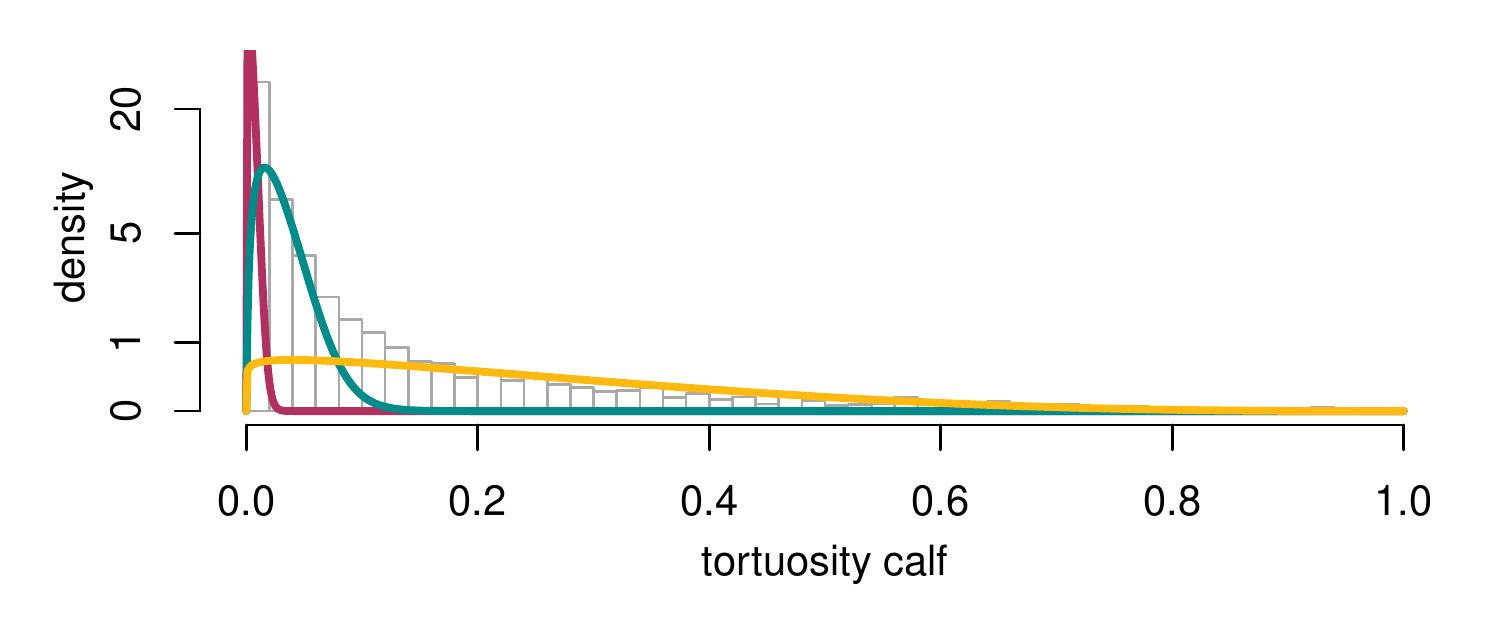}
   \caption{Estimated state-dependent distributions for tortuosity of the dolphin mother and calf, respectively, weighted by the stationary distribution of the bivariate Markov chain.}
    \label{fig:tortuosity_sdd_CHMM}
\end{figure}

\begin{figure}[t!]
    \centering
    \centering
    \includegraphics[width=0.7\textwidth]{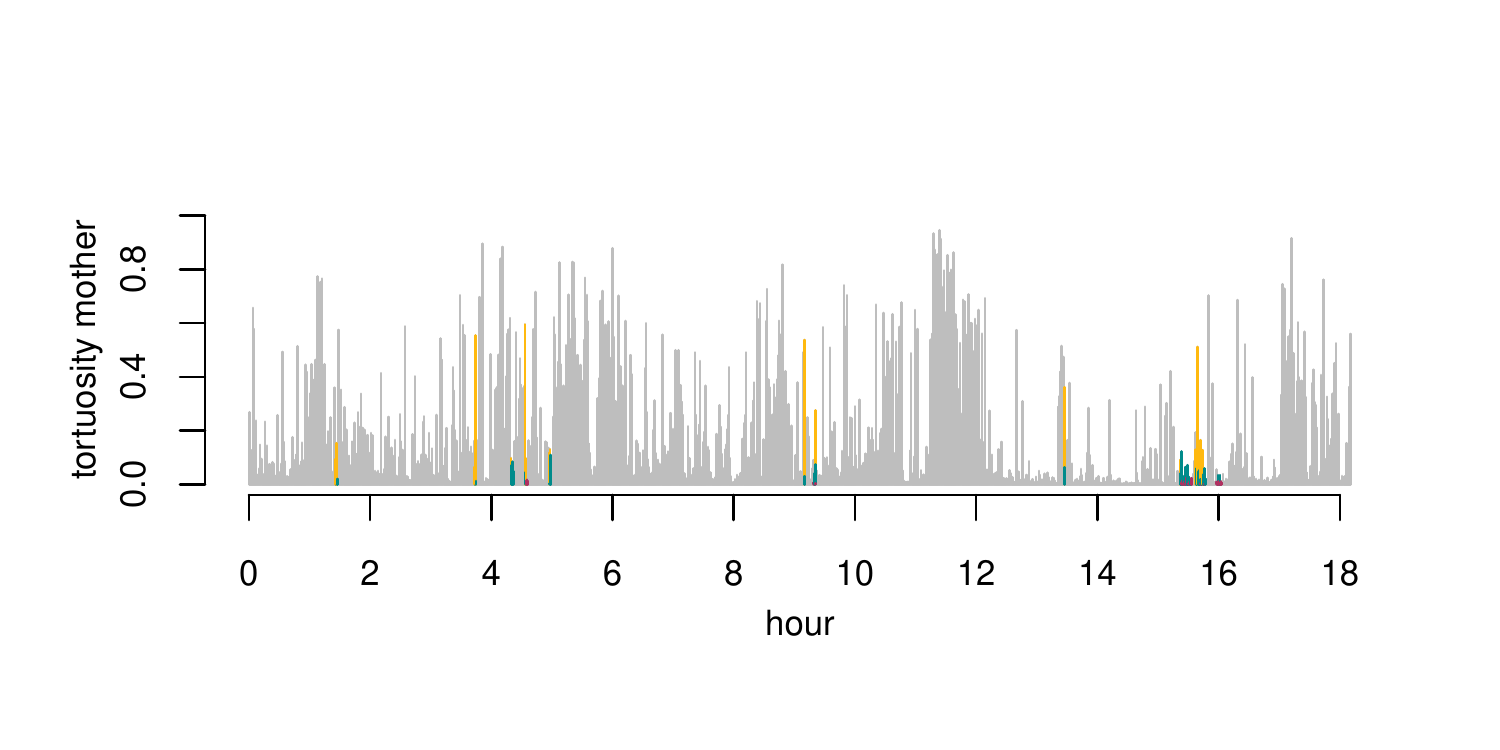}
    
    \vspace{-5em}
        
    \includegraphics[width=0.7\textwidth]{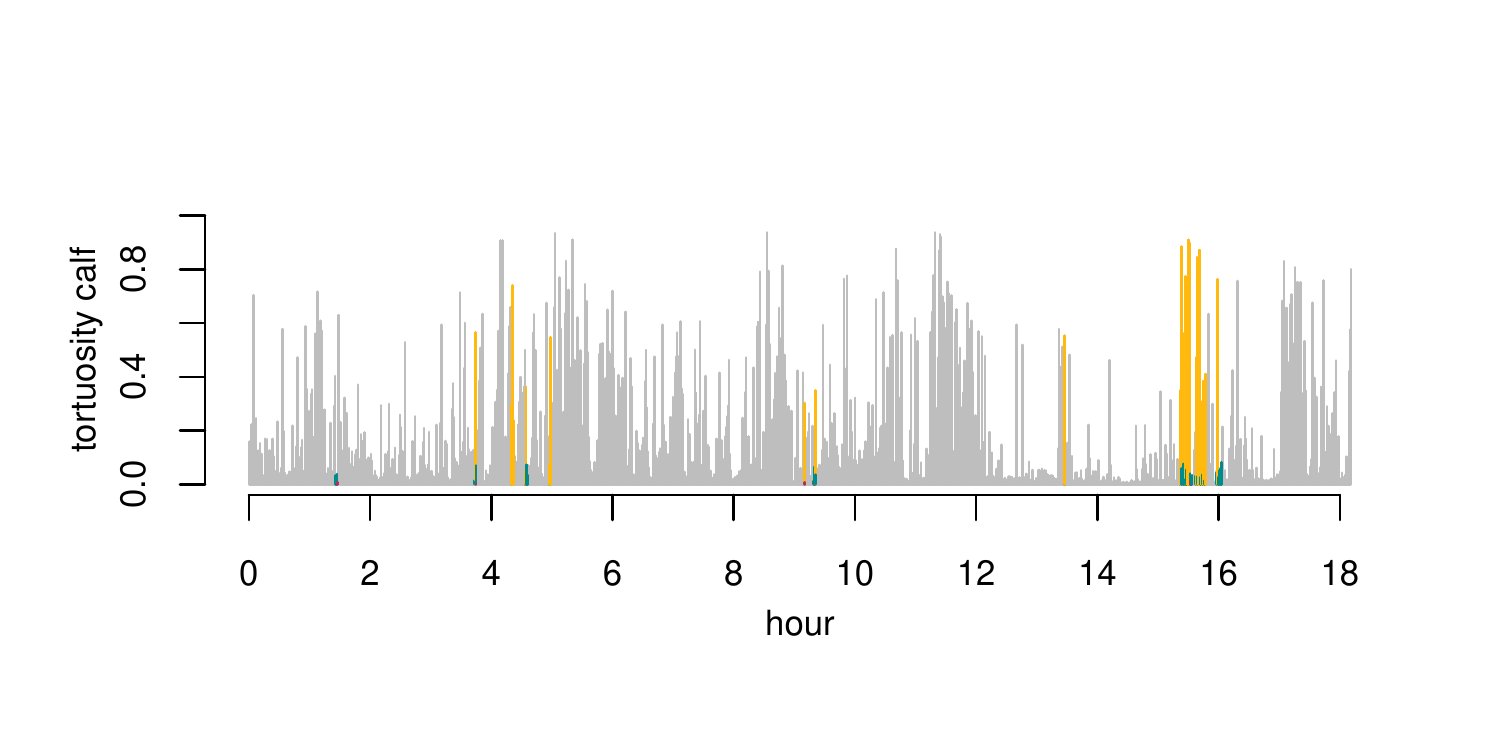}
    \vspace{-2em}
    \caption{Tortuosity time series of dolphin mother and calf, with states differing between mother and calf highlighted in colour.}
    \label{fig:tortuosity_decoded}
\end{figure}

The estimated state-dependent beta distributions are displayed in Figure \ref{fig:tortuosity_sdd_CHMM}. For both animals, the model identifies similar movement patterns, with state 1 capturing low tortuosity values (approximate straight-line movement; means $0.004$ and $0.005$ for mother and calf, respectively), state 2 accommodating any moderately large tortuosity values ($0.026$ and $0.029$), and state 3 associated with the most tortuous movements ($0.228$ and $0.231$). According to the fitted CHMM, the movement patterns evolve almost synchronously, with the bivariate states $(1,1)$, $(2,2)$ and $(3,3)$ clearly dominating the state process (Table \ref{tab:stat-dist_dolphin}). According to the Viterbi-decoded state sequence, the dolphins occupied different behavioural modes in only $4\%$ of all 10-second intervals considered. The corresponding observations are highlighted in Figure \ref{fig:tortuosity_decoded}, indicating the calf's movement to occasionally be more tortuous than the mother's movement towards the end of the time series (potentially related to the calf foraging independently of the mother). 
\begin{table}[t!]
    \centering
    \begin{tabular}{c|ccccccccc}
        state &  (1,1) & (1,2) & (1,3) &  (2,1) & (2,2) & (2,3) &  (3,1) & (3,2) & (3,3) \\\midrule
        probability & 0.339 & 0.013 & 0.002 & 0.004 & 0.404 & 0.015 & 0.001 & 0.004 & 0.218 \\
    \end{tabular}
    \caption{Steady-state (stationary) probabilities of the state process as implied by the estimated TPM.}
    \label{tab:stat-dist_dolphin}
\end{table}
The identification of such differences can be used as a starting point for further biological inference. For example, environmental covariates could be incorporated for further investigations into the role and the causes of different state combinations. Overall, the results suggest that the movement behaviour of mother and calf is well adapted to each other.

\subsection{Electronic health record data}\label{Sec4.2}

In our second case study, we analyse electronic health record data of patients hospitalised in the ICU of the Ronald Reagan UCLA medical center. We use a subset of the data also considered in \citet{ala18a} and \citet{ala18b}. ICU patients usually suffer from severe illnesses and injuries and are intensively observed by the nurses and physicians. However, as the patients undergo an increased risk, it is important to understand the progression of diseases and to identify early indications of a forthcoming deterioration. Modelling and analysing the physiological processes over time could help to detect critical developments early and support the decision-making of the physicians. State-switching time series models provide an intuitive and convenient framework for modelling the evolution of a system over time, and hence to quantify the risk of an impending deterioration of a patient's health state.

The data contain hourly measurements of four major vital signs: heart rate (in beats per minute, bpm), respiratory rate (in breaths per minute, bpm), systolic and diastolic blood pressure (in millimetre of mercury, mmHg). We did not consider diastolic blood pressure as it is strongly correlated with systolic blood pressure (Pearson correlation of 0.58). The data set further contains information about sex, age, admission type and location for each patient. The medical diagnosis, however, is omitted. In order to reduce the substantial patient heterogeneity caused by the underlying diseases, in this case study we consider only the patients who undergo dialysis, and restrict our analysis to patients with known sex and age who stayed in the ICU for more than $24$ hours. This results in a sample size of $T=110,964$ hourly observations from $696$ hospitalised patients (44\% female; age 17-89 with a median of 62; 1-80 days in ICU with a median of 4 days).

\begin{figure}[t!]
    \centering
    \includegraphics[width=0.65\textwidth]{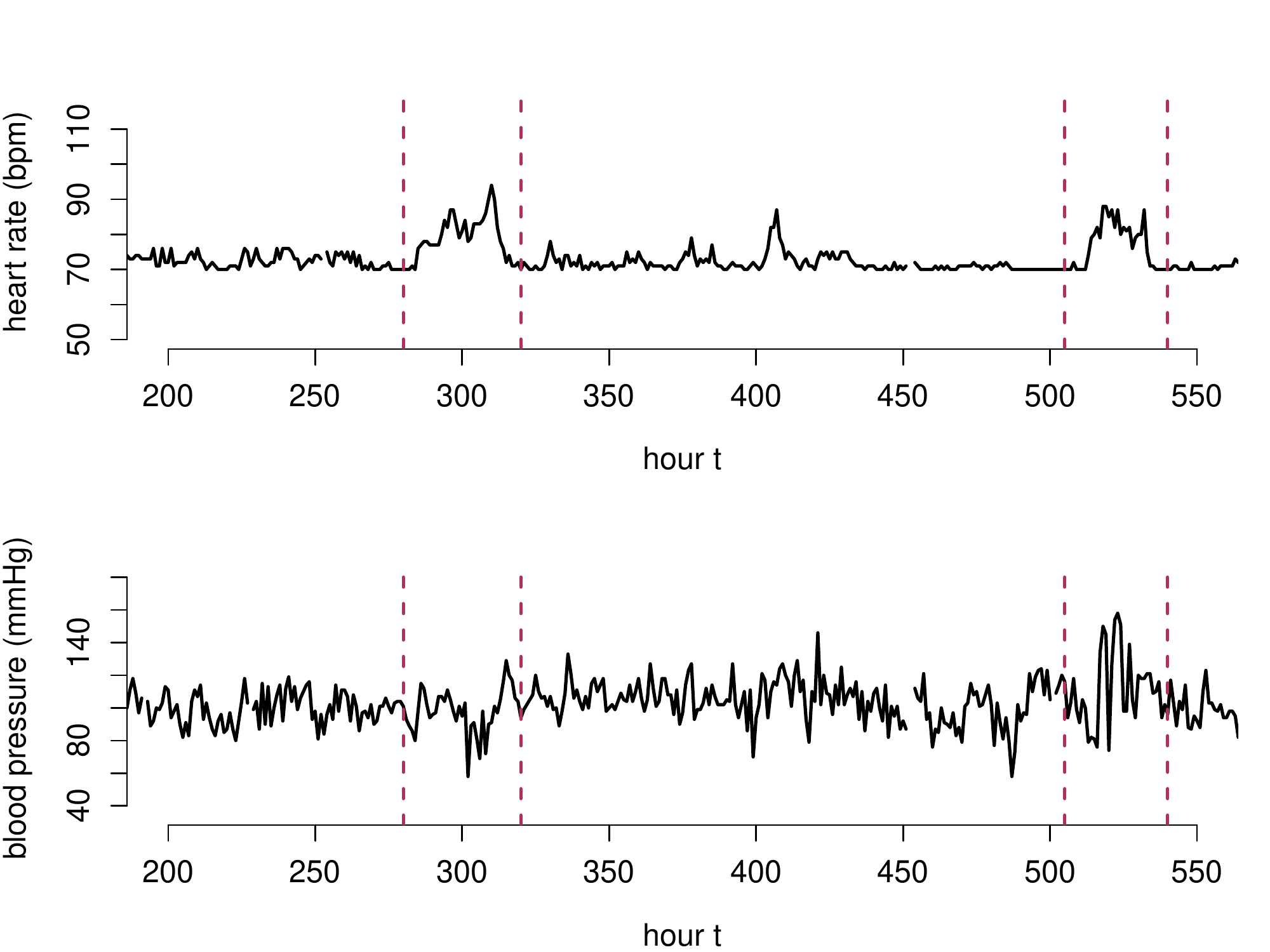}
    \caption{Example time series for heart rate and systolic blood pressure, respectively. The red lines highlight intervals with an elevated heart rate that does not seem in synchronity with the evolution of the observed systolic blood pressure.}
    \label{fig:HR_BP_example}
\end{figure}

The observed vital signs do not evolve synchronously over time --- for example, an increase in the heart rate is not necessarily accompanied by a change in blood pressure (cf.\ Figure \ref{fig:HR_BP_example}).
To account for such asynchronous evolution of the vital signs and the associated state of body functions, we consider a Cartesian product CMSR model with three states per vital sign, thus 27 state combinations in total. The model formulations with more restrictive dependence assumptions were again inferior in terms of the AIC (results not shown). All vital signs are modelled using state-dependent normal distributions, with the corresponding means additionally depending on the covariates sex and age:
$$\mu_{m,i}=\beta_{0,m,i}+\beta_{1,m,i} \cdot I_{\{\text{female}\}}+\beta_{2,m,i} \cdot \text{age},$$
for vital sign $m \in \{HR,RR,sBP\}$ and corresponding state $i=1,2,3$ (the patient index is omitted to simplify notation).
%Ruth: explain m and i

\begin{figure}[!tb]
    \centering
    \includegraphics[width=0.9\textwidth]{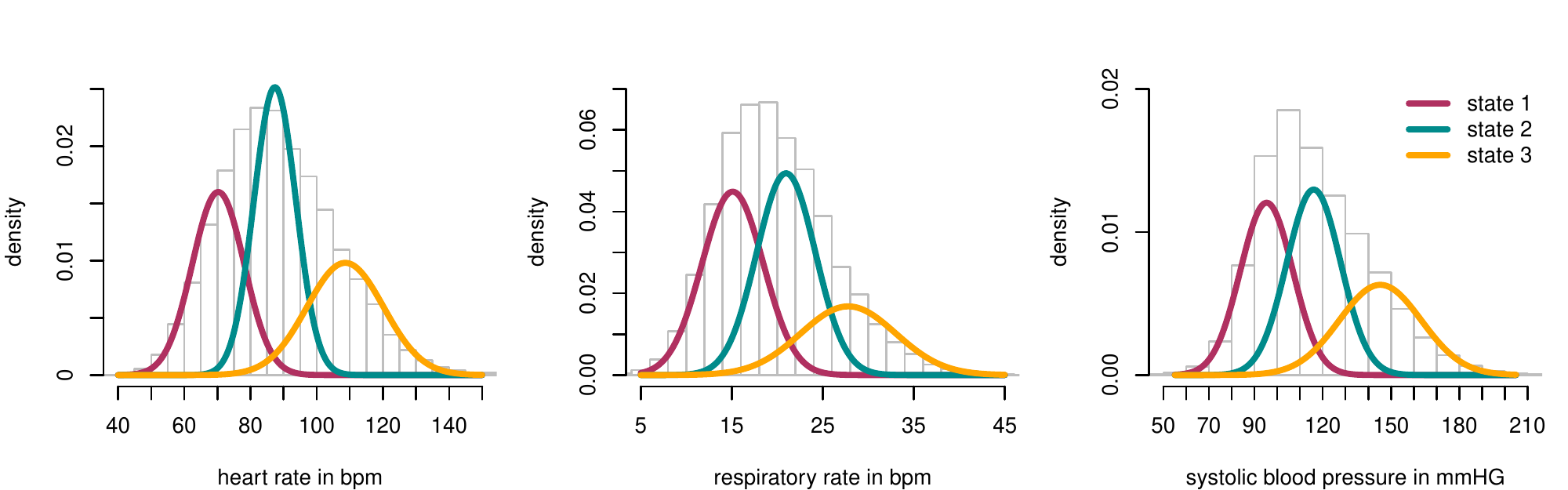}
    \caption{Estimated state-dependent distributions for heart rate, respiratory rate and systolic blood pressure, respectively, for 62-year-old males.}
    \label{fig:UCLA_CMSR3}
\end{figure}

Figure \ref{fig:UCLA_CMSR3} illustrates the estimated state-dependent distributions for male patients with the median age $62$. For each of the three vital signs, states 1, 2 and 3 effectively correspond to low, medium and high values, respectively. Some of the vital signs' underlying states allow for a direct interpretation: for example, the third systolic blood pressure state captures high values which may indicate some form of hypertension, the third respiratory rate state captures abnormally rapid breathing. However, for other states, the interpretation is less clear.

Table \ref{tab:UCLA_CMSR3} gives the estimates of the parameters associated with the state-dependent process, showing only small effects of the covariates considered. According to the model, we would expect to observe slightly lower heart rates, respiratory rates and systolic blood pressures for older patients. In case of respiratory rate and systolic blood pressure this is an unexpected result, which may be due to the exceptional circumstance of the patients considered being treated in the ICU. The estimated effects of the sex are relatively small. 
\begin{table}[!t]
    \centering
\begin{tabular}{lrrr}
\toprule[0.09 em]
                &    heart rate ($m=1$) &  resp.\ rate ($m=2$)   & blood press.\ ($m=3$)   \\\midrule[0.09 em]
$\beta_{0,m,1}$       &  70.40 (0.09)     &    15.09 (0.04)   &  95.50 (0.12)         \\[0.09 em]
$\beta_{1,m,1}$ (female)  &  -0.08 (0.12)     &    -0.09 (0.05)   &  0.31 (0.15)         \\[0.09 em]
$\beta_{2,m,1}$ (age)     &  -1.31 (0.06)     &   0.09 (0.03)     &   -1.60 (0.11)         \\[0.09 em]
$\sigma^2_{m,1}$          &    7.85 (0.03)    &    3.38 (0.02)    &   11.53 (0.05)         \\\midrule[0.09 em]
$\beta_{0,m,2}$       &   87.64 (0.08)    &     20.97 (0.05)  &  116.30 (0.15)        \\[0.09 em]
$\beta_{1,m,2}$ (female)  &  -0.09 (0.12)    &     -0.22 (0.06)  &   0.46 (0.13)       \\[0.09 em]
$\beta_{2,m,2}$ (age)     &  -2.57 (0.05)     &   -0.02 (0.04)    &  -2.51 (0.10)        \\[0.09 em]
$\sigma^2_{m,2}$          &    6.35 (0.03)    &    3.23 (0.02)    &   11.85 (0.06)       \\\midrule[0.09 em]
$\beta_{0,m,3}$       & 108.97 (0.12)    &     27.82 (0.07)  &  145.63 (0.20)        \\[0.09 em]
$\beta_{1,m,3}$ (female)  &  -0.62 (0.15)     &     0.18 $\thickspace$ (0.10)   &    1.87 (0.22)        \\[0.09 em]
$\beta_{2,m,3}$ (age)     & -3.42  (0.08)     &     -0.23 (0.04)  &   -3.28 (0.10)        \\[0.09 em]
$\sigma^2_{m,3}$          &  11.57 (0.05)     &     5.22 (0.03)   &   18.02 (0.09)       \\
\bottomrule[0.09 em]
\end{tabular}
    \caption{Estimated parameters (and standard errors) associated with the state-dependent distributions for heart rate, respiratory rate and systolic blood pressure, respectively.}
    \label{tab:UCLA_CMSR3}
\end{table}

\begin{figure}[!t]
    \centering
    \includegraphics[width=0.75\textwidth]{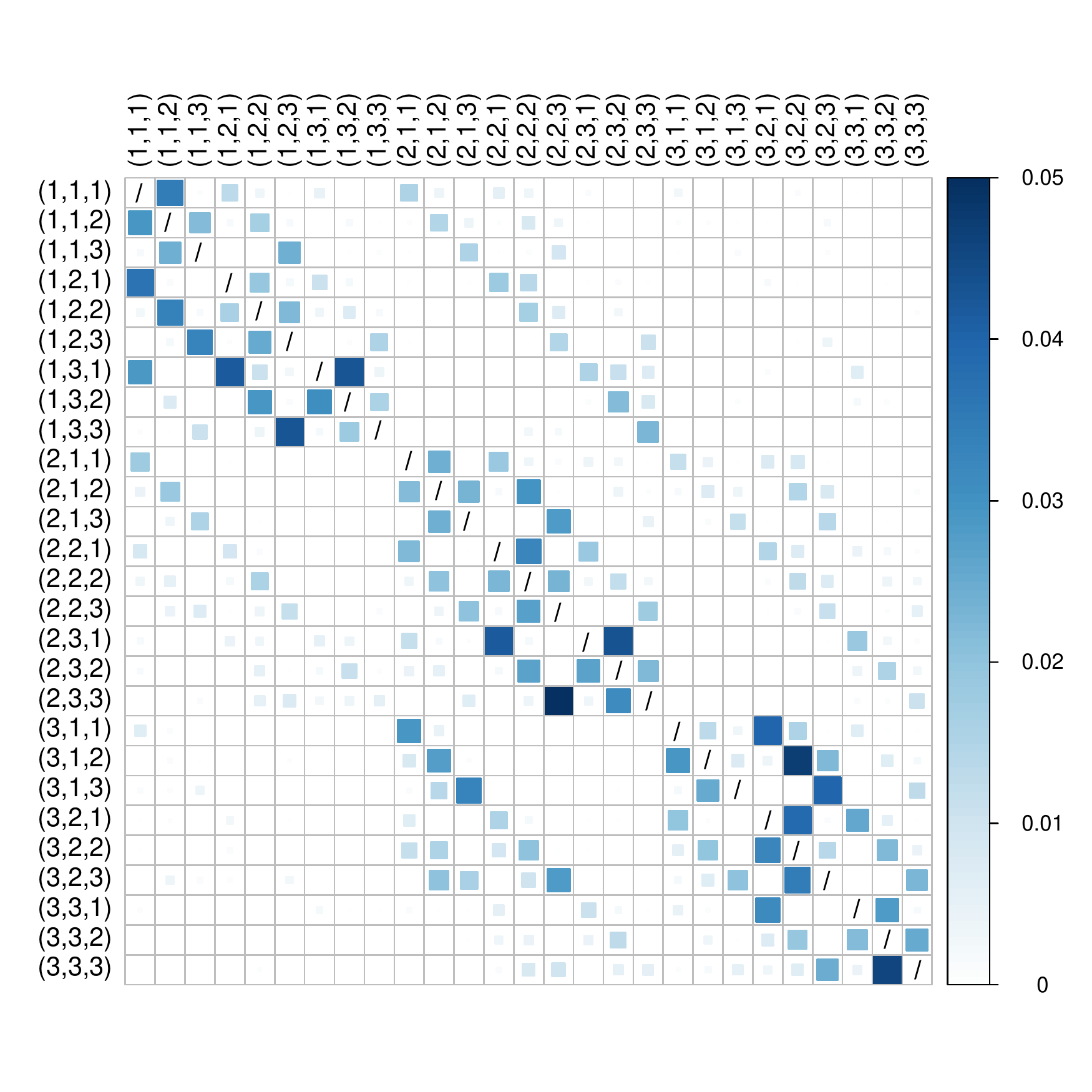}
    \caption{Off-diagonal elements of the estimated transition probability matrix. The diagonal entries lie between $0.830$ and $0.922$.}
    \label{fig:UCLA_gamma}
\end{figure}

The diagonal elements of the estimated $27 \times 27$ TPM, i.e.\ the probabilities to remain in the current state, lie between $0.830$ and $0.922$, indicating persistence in all bivariate states. The off-diagonal elements as displayed in Figure \ref{fig:UCLA_gamma} illustrate the estimated state dynamics. Most transition probabilities are estimated close to zero, with only infrequent abrupt switches from `low value' states to `high value' states. In some instances, the heart rate's state variable seems to dominate the process. For instance, given state vector $(1,2,2)$, the process is more likely to switch to state $(1,1,2)$ than $(2,2,2)$, and given state $(2,1,1)$, transitions to state $(2,1,2)$ or $(2,2,1)$ are more likely than a switch to $(1,1,1)$ --- these could be indications that the other state variables tend to adapt to the heart rate's state variable.  Overall, according to the stationary distribution, the most probable state combination is $(2,2,2)$, hence the `medium value' state for all vital signs.

The main advantage of the full Cartesian product CMSR model is that it allows us to derive a completely data-driven dependence structure of how the multivariate state process evolves over time. While our model is still somewhat simplistic, e.g.\ with regard to the conditional independence assumption, it offers an idea of the type of inference that can be gleaned on the joint evolution of heart rate, respiratory rate, and blood pressure. Such results could further be used, for example, to develop risk scores based on the probabilities to switch to deterioration states, or to cluster the different courses of diseases based on the patients' Viterbi sequences.

%%%%%%%%%%%%%%%%%%%%%%%%%%% 5: Discussion and Conclusions %%%%%%%%%%%%%%%%%%%%%%%%%%%
\section{Discussion}\label{Sec5}

CHMMs constitute a natural extension of basic HMMs to address scenarios with multiple time series whose underlying state processes interact. The explicit modelling of dependencies between the state variables can increase estimation accuracy, may decrease state classification error, and generally provides new opportunities for meaningful inference related to the correlation between processes. The potential of CHMMs has already been recognised in particular in engineering, where these models have been applied in various classification and signal processing tasks such as action recognition \citep{bra97b}, audio-visual speech recognition \citep{nef02}, bearing fault recognition \citep{zho16}, and EEG, ECG and PCG classification \citep{mic14,oli17}. Due to technological advances for example in animal tracking and in electronic health recordings (as illustrated in Section \ref{Sec4}), and generally the rapid growth in the amount of multi-stream data collected, we anticipate CHMMs to gain popularity also in other statistical modelling tasks such as forecasting or general inference on data-generating processes. In addition to the application areas showcased in the present paper, CHMMs could for example be useful to model the spread of infection in individual-based epidemic models \citep{tou19}, for exploiting dependencies between different economic markets in financial risk management \citep{cao19}, or to accommodate the spatio-temporal correlation of meteorological and geophysical time series \citep{sto19}.

\begin{figure}[!t]
    \centering
    \includegraphics[width=0.85\textwidth]{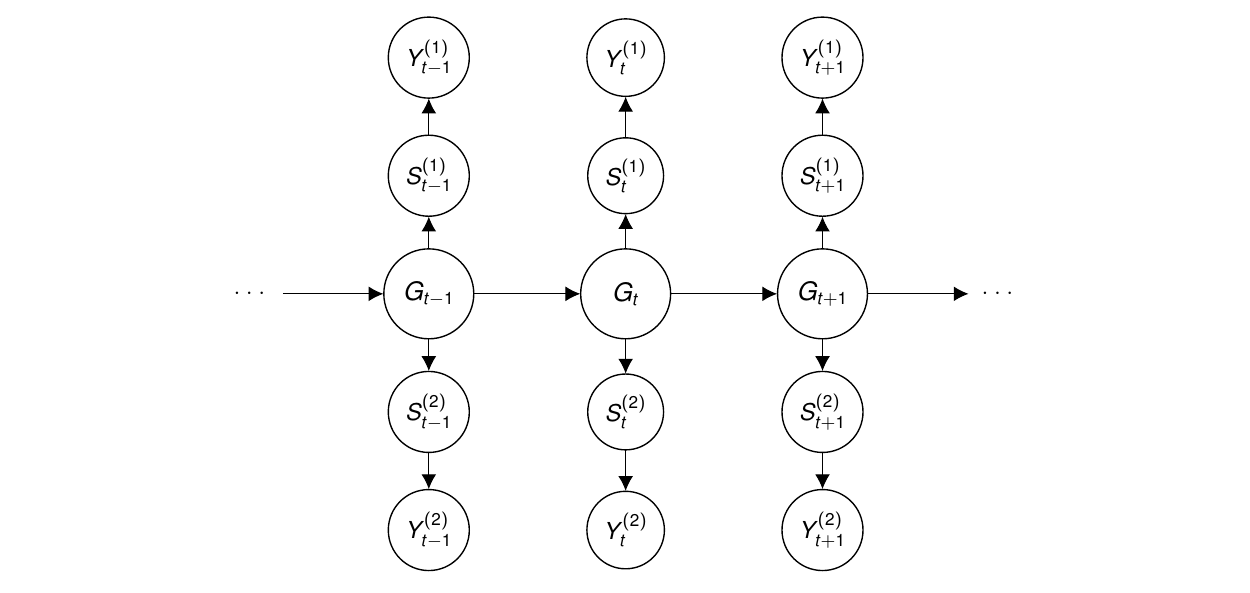}
    \caption{Possible hierarchical model with a global state $G_{t}$ determining the individual states $S_t^{(1)}$ and $S_t^{(2)}$, which in turn determine the distribution of the observed variables, $Y_{t}^{(1)}$ and $Y_{t}^{(2)}$.}
    \label{fig:discussion}
\end{figure}
The main barrier to CHMMs becoming much more widely used in applied statistics is the models' complexity arising from a curse of dimensionality: the number of model parameters very rapidly increases as the number of state variables or the number of states per variable increases, leading to high computational costs and numerical problems. Without imposing constraints on the model structure, CHMM-based analyses thus risk being limited to scenarios with only moderate numbers of variables and states. One possible way forward may be $\ell_1$ regularisation as suggested by \citet{bol17}, who use penalised estimation to arrive at a sparse dependence structure. We also expect alternative non-standard dependence structures for modelling interactions to become of increasing interest. For example, for interacting animals, it would be conceptually appealing (and mathematically convenient) to formulate models that are built around a global (``herd-level'') sequence of states $G_1,\ldots,G_T$, such that at any time $t$ the $M$ individual states $S_t^{(m)}$, $m=1,\ldots,M$, are drawn from a distribution determined by $G_t$ (see Figure \ref{fig:discussion} for an illustration with $M=2$). Such a model would not suffer from the curse of dimensionality, yet the global state process would still induce correlation between individuals, with the individuals' state processes occasionally deviating from the dominant group pattern (similar in spirit to, e.g., \citealp{zha05}, \citealp{lan14}). Mathematically, this model is simply an HMM with state-dependent mixture distributions, such that inference would be straightforward. The investigation of such alternative dependence structures as well as efficient and robust inferential approaches for conventional CHMMs are promising avenues for future research.

\section*{Acknowledgements}
We would like to thank Dr.\ Scott Hu (Division of Pulmonary and Critical Care Medicine, David Geen School of Medicine, UCLA) for providing the clinical data. We would also like to thank P. Tyack, L. Sayigh, V. Janik and R. Wells, the Chicago Zoological Society and Sarasota Dolphin Research Project staff scientists, and the funders and volunteers involved in Sarasota dolphin health assessments. Dolphin data collection was carried out under NMFS Research Permit No. 15543 to RSW and with IACUC approvals through Woods Hole Oceanographic Institution, Mote Marine Laboratory and the UK Animal Welfare and Ethics Committee. This work was supported by The Alan Turing Institute under the EPSRC Grant EP/N510129/1. RK was funded by a Leverhulme Research Fellowship.

\end{spacing}

\begin{thebibliography}{99}

\bibitem[\protect\citeauthoryear{Alaa {\em et~al.\/}}{2018}]{ala18b}
Alaa, A.M., Yoon, J., Hu, S.\ and van der Schaar, M.\ (2018).
Personalized risk scoring for critical care prognosis using mixtures of Gaussian processes.
\textit{IEEE Transactions on Biomedical Engineering}, \textbf{65}, 207--218.

\bibitem[\protect\citeauthoryear{Alaa and van der Schaar}{2018}]{ala18a}
Alaa, A.M.\ and van der Schaar, M.\ (2018).
A hidden absorbing semi-Markov model for informatively censored temporal data: Learning and inference.
\textit{Journal of Machine Learning Research}, \textbf{19}, 1--62.

\bibitem[\protect\citeauthoryear{Baker}{1975}]{dra75}
Baker, J.\ (1975).
The DRAGON system --- An overview.
\textit{IEEE Transactions on Acoustics, Speech, and Signal Processing}, \textbf{23}, 24--29.

\bibitem[\protect\citeauthoryear{Bolton {\em et~al.\/}}{2017}]{bol17}
Bolton, T., Tarun, A., Sterpenich, V., Schwartz, S.\ and De Ville, D.\ (2017).
Interactions between large-scale functional brain networks are captured by sparse coupled HMMs.
\textit{IEEE Transactions on Medical Imaging}, \textbf{37}, 230--240.

\bibitem[\protect\citeauthoryear{Brand}{1997}]{bra97a}
Brand, M.\ (1997).
Coupled hidden Markov models for modeling interacting processes.
\textit{Technical Report 405}, MIT Media Laboratory, Cambridge.

\bibitem[\protect\citeauthoryear{Brand {\em et~al.\/}}{1997}]{bra97b}
Brand, M., Olivier, N.\ and Pentland, A.\ (1997).
Coupled hidden Markov models for complex action recognition.
\textit{Proceedings of IEEE Computer Society Conference on Computer Vision and Pattern Recognition}, 994--999.

\bibitem[\protect\citeauthoryear{Brewer {\em et~al.\/}}{2006}]{bre06}
Brewer, N., Liu, N., De Vel, O.\ and Caelli, T.\ (2006).
Using coupled hidden Markov models to model suspect interactions in digital forensic analysis.
\textit{Proceedings of International Workshop on Integrating AI and Data Mining}, 58-–64.

\bibitem[\protect\citeauthoryear{Bulla and Bulla}{2006}]{bul06}
Bulla, J.\ and Bulla, I.\ (2006).
Stylized facts of financial time series and hidden semi-Markov models.
\textit{Computational Statistics \& Data Analysis}, \textbf{51}, 2192--2209.

\bibitem[\protect\citeauthoryear{Cao {\em et~al.\/}}{2019}]{cao19}
Cao, W., Zhu, W.\ and Demazeau, Y.\ (2019).
Multi-layer coupled hidden Markov model for cross-market behavior analysis and trend forecasting.
\textit{IEEE Access}, \textbf{7}, 158563--158574.

\bibitem[\protect\citeauthoryear{Ghahjaverestan {\em et~al.\/}}{2016}]{gha16}
Ghajaverestan, N.M., Masoudi, S., Shamsollahi, M.B., Beuchée, A., Plady, P., Ge, D.\ and Hern\'{a}ndez, A.I.\ (2016).
Coupled hidden Markov model-based method for apnea bradycardia detection.
\textit{IEEE Journal of Biomedical and Health Informatics}, \textbf{20}, 527--538.

\bibitem[\protect\citeauthoryear{Ghosh {\em et~al.\/}}{2017}]{gho17}
Ghosh, S., Li, J., Cao, L. and Ramamohanarao, K.\ (2017).
Septic shock prediction for ICU patients via coupled HMM walking
on sequential contrast patterns.
\textit{Journal of Biomedical Informatics}, \textbf{66}, 19--31.

\bibitem[\protect\citeauthoryear{Hamilton}{2008}]{ham08}
Hamilton, J.D.\ (2008), Regime-switching models. In Durlauf, S.N.\ and Blume, L.E., eds. \textit{The new palgrave dictionary of economics}, pages 5471--5475. Palgrave Macmillan, London.

\bibitem[\protect\citeauthoryear{Johnson {\em et~al.\/}}{2016}]{joh16}
Johnson, D.S., Laake, J.L., Melin, S.R.\ and DeLong, R.L.\ (2016).
Multivariate state hidden Markov models for mark-recapture data.
\textit{Statistical Science}, \textbf{31}, 233--244.

\bibitem[\protect\citeauthoryear{Langrock {\em et~al.\/}}{2014}]{lan14}
Langrock, R., Hopcraft, J.G.C., Blackwell, P.G., Goodall, V., King, R., Niu, M., Patterson, T.A., Perdersen, M.W., Skarin, A.\ and Schick, R.S.\ (2014).
Modelling group dynamic animal movement.
\textit{Methods in Ecology and Evolution}, \textbf{5}, 190--199.

\bibitem[\protect\citeauthoryear{Langrock {\em et~al.\/}}{2012}]{lan12}
Langrock, R., King, R., Matthiopoulos, J., Thomas, L., Fortin, D.\ and Morales, J.M. (2012).
Flexible and practical modeling of animal telemetry data: hidden Markov models and extensions.
\textit{Ecology}, \textbf{93}, 2336--2342.

\bibitem[\protect\citeauthoryear{Langrock {\em et~al.\/}}{2017}]{lan17}
Langrock, R., Kneib, T., Glennie, R.\ and Michelot, T.\ (2017).
Markov-switching generalized additive models. 
\textit{Statistics and Computing}, \textbf{27}, 259–-270.

\bibitem[\protect\citeauthoryear{Langrock {\em et~al.\/}}{2013}]{lan13}
Langrock, R., Swihart, B.\ J., Caffo, B.\ S., Crainiceanu, C.\ M.\ and Punjabi, N.\ M. (2013).
Combining hidden Markov models for comparing the dynamics of multiple sleep electroencephalograms.
\textit{Statistics in Medicine}, \textbf{32}, 3342--3356.

\bibitem[\protect\citeauthoryear{Lin {\em et~al.\/}}{2012}]{lin12}
Lin, J., Wu, C.\ and Wei, W.\ (2012).
Error weighted semi-coupled hidden Markov model for audio-visual emotion recognition.
\textit{IEEE Transactions on Multimedia}, \textbf{14}, 142--156.

\bibitem[\protect\citeauthoryear{Maruotti {\em et~al.\/}}{2019}]{mar19}
Maruotti, A., Punzo, A.\ and Bagnato, L.\ (2019).
Hidden Markov and semi-Markov models with multivariate leptokurtic-normal components for robust modeling of daily returns series.
\textit{Journal of Financial Econometrics}, \textbf{17}, 91–-117. 

\bibitem[\protect\citeauthoryear{Michalopoulos and  Bourbakis}{2014}]{mic14}
Michalopoulos, K.\ and Bourbakis, N.\ (2014).
Using dynamic Bayesian networks for modeling EEG topographic sequences.
\textit{36th Annual International Conference of the IEEE Engineering in Medicine and Biology Society}, \textbf{7}, 4928--4931.

\bibitem[\protect\citeauthoryear{Michelot {\em et~al.\/}}{2016}]{mic16}
Michelot, T., Langrock, R.\ and Patterson, T.A.\ (2016).
moveHMM: An R package for analysing animal movement data using hidden Markov models.
\textit{Methods in Ecology and Evolution}, \textbf{7}, 1308--1315.

\bibitem[\protect\citeauthoryear{Nefian {\em et~al.\/}}{2002}]{nef02}
Nefian, A.V., Liang, L., Pi, X., Liu, X.\ and Murphy, K.\ (2002). 
Dynamic Bayesian networks for audio-visual speech recognition. \textit{EURASIP Journal on Advances in Signal Processing}, \textbf{11}, 1274–-1288.

\bibitem[\protect\citeauthoryear{Oliveira {\em et~al.\/}}{2002}]{oli17}
Oliveira, J., Sousa, C.\ and Coimbra, M.T.\ (2002). Coupled hidden Markov model for automatic ECG and PCG segmentation.
\textit{IEEE International Conference on Acoustics, Speech and Signal Processing}, 1023--1027.

\bibitem[\protect\citeauthoryear{Raftery}{1985}]{raf85}
Raftery, A.\ (1985).
A model for high-order Markov chains.
\textit{Journal of the Royal Statistical Society B}, \textbf{47}, 528--539.

\bibitem[\protect\citeauthoryear{Rezek {\em et~al.\/}}{2000}]{rez00}
Rezek, I., Sykacek, P.\ and Roberts, S.J.\ (2000).
Learning interaction dynamics with coupled hidden Markov models.
\textit{IEE Proceedings - Science, Measurement and Technology}, \textbf{147}, 345--350. Erratum in \textit{IEE Proceedings - Science, Measurement and Technology}, \textbf{148}, 221.

\bibitem[\protect\citeauthoryear{Ryd\'{e}n}{2008}]{ryd08}
Ryd\'{e}n, T.\ (2008).
EM versus Markov chain Monte Carlo for estimation of hidden Markov models: A computational perspective.
\textit{Bayesian Analysis}, \textbf{3}, 659--688.

\bibitem[\protect\citeauthoryear{Saul and Jordan}{1999}]{sau99}
Saul, L.K.\ and Jordan, M.I.\ (1999).
Mixed memory Markov models: Decomposing complex stochastic processes as mixtures of simpler ones.
\textit{Machine Learning}, \textbf{37}, 75–-87.

\bibitem[\protect\citeauthoryear{Sherlock {\em et~al.\/}}{2013}]{she13}
Sherlock, C., Xifara, T., Telfer, S.\ and Begon, M.\ (2013).
A coupled hidden Markov model for disease interactions.
\textit{Journal of the Royal Statistical Society, Series C}, \textbf{62}, 609--627.

\bibitem[\protect\citeauthoryear{Stoner and Economou}{2019}]{sto19}
Stoner, O.\ and Economou, T.\ (2019).
A comprehensive hidden Markov model for hourly rainfall time series.
\textit{arXiv}, arXiv:1906.03846.

\bibitem[\protect\citeauthoryear{Touloupou {\em et~al.\/}}{in press}]{tou19}
Touloupou, P., Finkenstädt, B.\ and Spencer, S.E.F.\ (in press).
Scalable Bayesian inference for coupled hidden Markov and semi-Markov models.
\textit{Journal of Computational and Graphical Statistics}, DOI:10.1080/10618600.2019.1654880.

\bibitem[\protect\citeauthoryear{Visser {\em et~al.\/}}{2002}]{vis02}
Visser, I., Raijmakers, M.E.J.\ and Molenaar, P.\ (2002).
Fitting hidden Markov models to psychological data.
\textit{Scientific Programming}, \textbf{10}, 185--199.

\bibitem[\protect\citeauthoryear{Zhang {\em et~al.\/}}{2006}]{zha05}
Zhang, D., Gatica-Perez, D., Bengio, S.\ and Roy, D.\ (2006). 
Learning influence among interacting Markov chains.
\textit{Advances in Neural Information Processing Systems}, \textbf{61}, 1577--1584.

\bibitem[\protect\citeauthoryear{Zhou {\em et~al.\/}}{2016}]{zho16}
Zhou, H., Chen, J., Dong, G., Wang, H., Yuan, H.\ (2016). 
Bearing fault recognition method based on neighbourhood component analysis and coupled hidden Markov model.
\textit{Mechanical Systems and Signal Processing}, \textbf{66-67}, 568--581.

\bibitem[\protect\citeauthoryear{Zucchini {\em et~al.\/}}{2016}]{zuc16}
Zucchini, W., MacDonald, I.L.\ and Langrock, R.\ (2016).
{\em Hidden Markov models for time series: An introduction using R}. 
Second Edition, Chapman and Hall/CRC, Boca Raton.


\end{thebibliography}
\end{document}